%% file: main.tex
\documentclass{article}
\usepackage[T1]{fontenc} 
\usepackage[utf8]{inputenc} 
\usepackage{ismir,amsmath,cite,url}
\usepackage{graphicx}

\usepackage{subcaption}

\usepackage{amssymb}
\usepackage{pifont}
\usepackage{multirow}
\usepackage[symbol]{footmisc}
\usepackage{booktabs}
\usepackage[table,xcdraw]{xcolor}
\usepackage{import}

\definecolor{lightgray}{RGB}{230,230,230}
\newcommand{\grayline}{\arrayrulecolor{lightgray}\hline\arrayrulecolor{black}}

\usepackage{dsfont}

\usepackage{lineno}

\title{Semi-supervised Contrastive Learning of Musical Representations}

\multauthor
{Julien Guinot$^{1,2}$ \hspace{1cm} Elio Quinton$^2$ \hspace{1cm} György Fazekas$^1$} { 
$^1$ Centre for Digital Music, Queen Mary University of London, U.K.\\
$^2$ Music \& Audio Machine Learning Lab, Universal Music Group, London, U.K.\\
{\tt\small j.guinot@qmul.ac.uk}
}

\def\authorname{Julien Guinot, Elio Quinton, György Fazekas}

\usepackage[bookmarks=false,pdfauthor={\authorname},pdfsubject={\papersubject},hidelinks]{hyperref}

\hypersetup{
    colorlinks=true,
    linkcolor=blue,
    citecolor=blue,
    filecolor=magenta,      
    urlcolor=cyan,
    pdftitle={Overleaf Example},
    pdfpagemode=FullScreen,
    }

\begin{document}

\maketitle
\begin{abstract}


Despite the success of contrastive learning in Music Information Retrieval, the inherent ambiguity of contrastive self-supervision presents a challenge. Relying solely on augmentation chains and self-supervised positive sampling strategies can lead to a pretraining objective that does not capture key musical information for downstream tasks. We introduce semi-supervised contrastive learning (SemiSupCon), a simple method for leveraging musically informed labeled data (supervision signals) in the contrastive learning of musical representations. Our approach introduces musically relevant supervision signals into self-supervised contrastive learning by combining supervised and self-supervised contrastive objectives in a simpler framework than previous approaches. This framework improves downstream performance and robustness to audio corruptions on a range of downstream MIR tasks with moderate amounts of labeled data. Our approach enables shaping the learned similarity metric through the choice of labeled data that (1) infuses the representations with musical domain knowledge and (2) improves out-of-domain performance with minimal general downstream performance loss. We show strong transfer learning performance on musically related yet not trivially similar tasks - such as pitch and key estimation. Additionally, our approach shows performance improvement on automatic tagging over self-supervised approaches with only 5\% of available labels included in pretraining.

\end{abstract}

\section{Introduction}
\label{sec:intro}


Self-supervised learning (SSL) has emerged as a powerful paradigm for learning structured representations of data without the need for costly and time-consuming labeling. SSL approaches have achieved competitive performance on downstream tasks with minimal labeled data in many domains \cite{chenSimpleFrameworkContrastive2020, oordRepresentationLearningContrastive2019, grillBootstrapYourOwn2020, chenExploringSimpleSiamese2020, liMERTAcousticMusic2023, gongSSASTSelfSupervisedAudio2022, hsuHuBERTSelfSupervisedSpeech2021, spijkervetContrastiveLearningMusical2021}. In the field of Music Information Retrieval (MIR), the complexity of labeling for many tasks - due to the high technicality and subjectivity - underscores the importance of such self-supervised methods \cite{spijkervetContrastiveLearningMusical2021,vasquezTAILEDUNETMULTISCALE2022,zhaoS3TSelfSupervisedPretraining2022,liMERTAcousticMusic2023, liMAPMusic2VecSimpleEffective2022, wonFoundationModelMusic2023, mccallumSupervisedUnsupervisedLearning2022, castellonCodifiedAudioLanguage2021}. Instance-discriminative SSL specifically, such as contrastive learning, has proven to be effective in learning meaningful representations for a multitude of downstream tasks \cite{spijkervetContrastiveLearningMusical2021,vasquezTAILEDUNETMULTISCALE2022, mancoContrastiveAudioLanguageLearning2022}. However, major design choices such as positive mining strategies and augmentations are crucial to downstream performance \cite{xiao2021what,tianWhatMakesGood2020,spijkervetContrastiveLearningMusical2021,choiProperContrastiveSelfsupervised2022, mccallum2024effect}, and selecting a strategy for a given task remains a challenge, prompting the reintroduction of supervision within the SSL framework. In MIR, the key notion of ``similarity'' in contrastive learning can derive from a variety of musical attributes.  Guiding the model towards a musically informed similarity metric is an objective that may be achieved by leveraging supervised labeled data, i.e. \emph{supervision signals}.

In this work, we propose a novel semi-supervised contrastive learning method, SemiSupCon. Our method leverages both unlabeled and labeled data for contrastive learning, an extension of Contrastive Learning of Musical Representations (CLMR) in the music domain \cite{spijkervetContrastiveLearningMusical2021} and SupCon \cite{khosla2020supervised} in Computer Vision. Our approach differs from previous attempts at combining self-supervised contrastive learning with an auxiliary supervision signal in that it is the first to our knowledge to implement a fully-contrastive semi-supervised learning pipeline. The simple machinery of this method allows for leveraging new supervision signals beyond labels within the contrastive objective.

Briefly, the contributions of this work are the following: (1) We propose an architecturally simple extension of self-supervised and supervised contrastive learning to the semi-supervised case with the ability to make use of a variety of supervision signals. (2) We show the ability of our method to shape the representations according to the support supervision signal used for the learning task with minimal performance loss on other tasks. (3) We propose a representation learning framework with low-data regime potential and higher robustness to data corruption. Our implementation and experiments are made publicly available at \hyperlink{https://github.com/Pliploop/SemiSupCon}{https://github.com/Pliploop/SemiSupCon}

\section{Related work}
\label{sec : Background}

Self-supervised learning aims to learn representations that capture the semantic structure of data without labels in order to utilize these representations on downstream tasks. Among self-supervised learning approaches, Contrastive Learning teaches a model to identify augmented samples originating from the same data point amongst distractor negative samples \cite{chenSimpleFrameworkContrastive2020, spijkervetContrastiveLearningMusical2021}. Beyond its success in neighboring fields, MIR and audio representation learning have largely benefited from Contrastive Learning approaches \cite{spijkervetContrastiveLearningMusical2021,vasquezTAILEDUNETMULTISCALE2022, al-tahanCLARContrastiveLearning2020a, chungW2vBERTCombiningContrastive2021,oordRepresentationLearningContrastive2019, fonsecaUnsupervisedContrastiveLearning2020CLSER}. From the implementation of CLMR, several works have expanded on contrastive learning for music, with competitive results on many downstream tasks and in multiple modalities \cite{mccallumSupervisedUnsupervisedLearning2022, zhaoS3TSelfSupervisedPretraining2022, vasquezTAILEDUNETMULTISCALE2022, yaoContrastiveLearningPositiveNegative2022, garoufisMultiSourceContrastiveLearning2023}. One of the key challenges of contrastive learning is establishing an effective positive mining strategy to select positive and negative samples\cite{tianWhatMakesGood2020,xiao2021what, choiProperContrastiveSelfsupervised2022}. Previous studies show that both the positive mining strategy and the augmentation chain are crucial toward the performance on a given downstream task \cite{xiao2021what,tianWhatMakesGood2020,mccallum2024effect,choiProperContrastiveSelfsupervised2022} - an inappropriate sampling strategy can lead to treating similar samples as negatives, to the detriment of downstream performance \cite{guoUltimateNegativeSampling2023,geRobustContrastiveLearning2021,huynhBoostingContrastiveSelfSupervised2022}. In MIR specifically, even the temporal proximity of two positive segments within an audio clip is influential on downstream performance depending on the task, as shown in \cite{choiProperContrastiveSelfsupervised2022}. Previous works have attempted to design domain-appropriate strategies for music and audio contrastive learning, including auxiliary similarity metrics \cite{akamaAuxiliarySelfsupervisionMetric2023,yaoContrastiveLearningPositiveNegative2022,manochaCDPAMContrastiveLearning2021, alonso2023pre}, weak supervision \cite{favoryCOALACoAlignedAutoencoders2020, mancoContrastiveAudioLanguageLearning2022, ferraroEnrichedMusicRepresentations2021, huangMuLanJointEmbedding2022}, as well as music-specific preprocessing and augmentations \cite{zhaoS3TSelfSupervisedPretraining2022, spijkervetContrastiveLearningMusical2021, garoufisMultiSourceContrastiveLearning2023}. 

Self-supervision is inherently limited by the ability of the positive mining strategy to select semantically relevant positives. Some approaches have attempted to reintroduce supervision signals for positive mining within the contrastive objective to reduce noise induced by self-supervised pseudolabels. SupCon \cite{khosla2020supervised} introduces supervised contrastive learning, which uses class labels to mine positives. Other approaches have extended contrastive learning to the semi-supervised regime by leveraging both labeled and unlabeled data. However, these approaches often use complex machinery, such as auxiliary classification modules or multiple losses \cite{akamaAuxiliarySelfsupervisionMetric2023, kim2021selfmatch, zhang2022semi, yang2022class}, making them inflexible and difficult to balance with regard to the supervision signal. Recently, in MIR, \textit{Akama et. al} \cite{akamaAuxiliarySelfsupervisionMetric2023} employ contrastive learning as an auxiliary loss for automatic tagging, with improved results over supervision alone.

\vspace{-10pt}

\section{Methods}

\subsection{Self-Supervised contrastive learning}
\label{subsec: SSL}

In the SSL setting for contrastive learning \cite{spijkervetContrastiveLearningMusical2021,chenSimpleFrameworkContrastive2020}, each sample in a $N$-sample batch is augmented into two views through a stochastic augmentation chain. Let $B$ be a batch of these augmented views $x_i$. Indices $i \in I = \{1,2...2N\}$ represent the index of a data point in the batch (anchor). $p(i)$ is the index of the augmented data point originating from the same original sample as the anchor (positive sample). $N(i)$ is the set of negatives: data points in the augmented batch excluding the anchor and positives: $N(i) = I \setminus \{i,p(i)\}$. Let $z_i$ be the embedded representation of the data point by an encoder $E : x \mapsto E(x) \in \mathbb{R}^{d_E}$ and a projection head $g : E(x) \mapsto g(E(x)) = z_i \in \mathbb{R}^{d_g}$ into the contrastive latent space. In the SSL setting, the objective function for the contrastive method is the normalised temperature-scaled cross-entropy loss \cite{chenSimpleFrameworkContrastive2020} between samples $i$ and $p(i)$ for all pairs in the batch:

\begin{equation}
    \mathcal{L}_{ssl}^i = - \log \frac{\exp(sim(z_i,z_{p(i)})/\tau)}{\sum\limits_{n \in N(i) \cup \{p(i)\}} \exp(sim(z_i,z_n/\tau))}
    \label{sslloss}
\end{equation}

Where $\tau$ is a temperature hyperparameter, $sim$ is a similarity function - usually, cosine similarity \cite{spijkervetContrastiveLearningMusical2021,chenSimpleFrameworkContrastive2020}. For the sake of brevity we notate $\sigma_{i,j} = \exp(sim(z_i,z_j)/\tau)$ in the rest of this work. 





\subsection{Supervised contrastive learning}

In the supervised setting \cite{khosla2020supervised}, The  set of \textit{supervised} positives $P_s(i)$ are now defined by the label information in the set of labels $y_i$: $ P_s(i) = \{p \in I | y_p = y_i\} \setminus i$. As in \cite{khosla2020supervised}, the supervised contrastive loss objective is given by:

\begin{equation}
    \mathcal{L}_{sl}^i  = \frac{-1}{|P_s(i)|}\sum\limits_{p\in P_s(i)} \log \frac{\sigma_{i,p}}{\sum\limits_{n\in N(i) \cup P_s(i)}\sigma_{i,n}}
\end{equation}

The contrastive matrix $\textbf{M}$ is constructed by leveraging class information obtained by mining the labels, i.e. if two samples $x_i$ and $x_j$ are in the same category then $\textbf{M}_{i,j} = 1$.




\subsection{Semi-supervised Contrastive Learning}
\label{sec: SMSL}

Let $\mathcal{U}$ be a set of unlabeled samples, and $\mathcal{S}^*$ be a set of labeled samples. We sample a proportion $p_s$ of the labeled dataset for training such that $|\mathcal{S}| = p_s|\mathcal{S}^*|$. Let $\mathcal{A} = \mathcal{U} \cup \mathcal{S}$ be the set of all data points seen during training. During training, we use both labeled and unlabeled data points by sampling batches $B$ comprised of proportions $b_s$ (resp. $1-b_s$) of labeled (resp. unlabeled) samples. $P_s(i) = \varnothing$ if $i$ is the index of an unlabeled data point.  We now define our semi-supervised contrastive loss, with $P_A(i) = P_s(i) \cup P_u(i)$, where $P_u(i)$ is the set of self-supervised positives \big($\{p(i)\}$ in Eq. \ref{sslloss}\big).:


\begin{equation}
    \resizebox{0.42\textwidth}{!}{
    $\mathcal{L}_{sem}^i = \frac{-1}{|P_A(i)|} \sum\limits_{p \in P_A(i)}  \log \Bigg( \frac{ \sigma_{i,p}}{\sum\limits_{n \in N(i) \cup P_A(i)} \sigma_{i,n}} \Bigg)$}
\label{semisupconloss}
\end{equation}

\begin{figure*}[h]
    \centering
    \includegraphics[width = \textwidth]{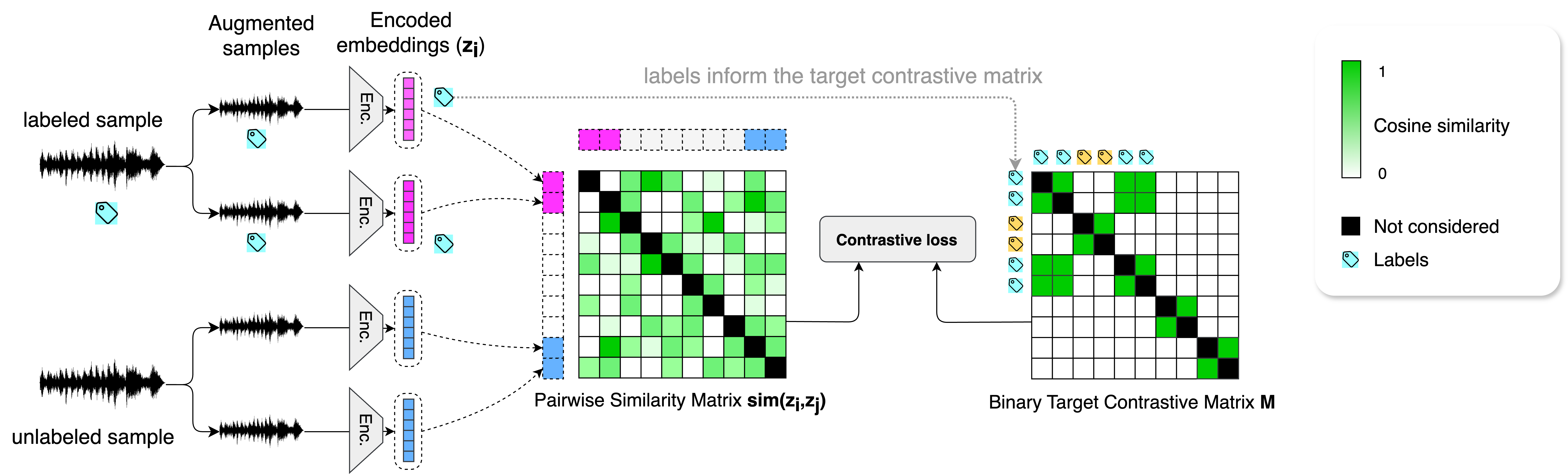}
    \caption{\small Semi-Supervised Contrastive Learning. The sparsely labeled dataset contains a mix of unlabeled data and labeled data. Given a batch, available labels (blue and yellow tags) are used to augment $\textbf{M}$. Unlabeled samples degenerate back to the self-supervised case. Loss is computed between the pairwise similarity matrix from the encoded embeddings and the target matrix using Equation \ref{semisupconloss}}
    \label{fig:SMSL}
\end{figure*}

 With the inclusion of both sets of positives, we generalize to both labeled and unlabeled data in our representation learning task: Note that if $\mathcal{U} = \varnothing$ or $\mathcal{S} = \varnothing$, the semi-supervised contrastive loss reverts back to the fully-supervised loss  or the fully self-supervised loss (as $P_s(i) = \varnothing$) respectively. The approach is shown Figure \ref{fig:SMSL}.

This approach differs from simply adding the supervised and self-supervised contrastive losses together, as our objective maintains the number of samples to discriminate against in the self-supervised setting by leveraging labeled data as negatives for the self-supervised samples.







\subsubsection{Extension to other supervision signals}
\label{sec: SMSL_sup_signals}

The range of supervision signals this method can leverage are limited only by the ability to construct the target contrastive matrix. In this, SemiSupCon can make use of support data beyond single label multiclass tasks. To demonstrate this, we devise two strategies for training on MagnaTagATune \cite{law2009evaluation}, which are studied in Section \ref{ablation: strategy}. For a multi-label signal, if $C \in \mathbb{N}$ labels coincide between two samples, we set the the corresponding index in the target contrastive matrix $\textbf{M}_{i,j} = 1$. the criterion $C$ is a hyperparameter which is studied in Section \ref{ablation: strategy}. By default we use $C=1$, i.e., if any labels coincide between two samples they are considered as positives.

Further, we can construct a target continuous similarity metric factor $\alpha_{i,j}$ which denotes the degree of ``semantic similarity`` between the samples by weighing the common classes by the total number of labels:

\begin{equation*}
    \alpha_{i,j} = \frac{2C_{i,j}}{(C_i + C_j)} 
\end{equation*}
    $C_{i,j}$ is the number of common classes for $x_i$ and $x_j$, $C_i$ and $C_j$ are the number of classes of $x_i$ and $x_j$. The similarity term $\sigma_{i,j}$ is then weighted by $\alpha_{i,j}$ in Eq. \ref{semisupconloss}.

\section{Experiments and results}


\subsection{Datasets}
\label{datasets}

For our experiments, we use The Free Music Archive (FMA) dataset \cite{fma_dataset} as a self-supervised dataset, i.e., we do not use its labels. To match the scale of the supervised datasets, we elect to use the \textit{medium} subset, containing 25000 clips of 30 seconds of audio.

We utilize several labeled datasets as support labeled data for training and for evaluation to demonstrate the cross-domain usefulness of SemiSupCon. For automatic tagging and most of our experiments, we use MagnaTagATune (MTAT) \cite{law2009evaluation} as labeled data as a proxy evaluation of general music understanding. We reproduce the canonical 12:3:1 train-test-validation splits \cite{spijkervetContrastiveLearningMusical2021}. We use MTG-Jamendo (all subsets, including the top 50 tags, genre, mood/theme, and instrument) \cite{bogdanov2019mtg} as another tagging dataset. We use NSynth \cite{nsynth2017} for pitch and instrument classificiation of short snippets, and MedleyDB \cite{bittner2014medleydb,bittner2016medleydb} for instrument classification with longer audio clips than NSyth. We use Giantsteps \cite{knees2015two} as a key classification dataset - as in \cite{yuan2024marble}, we use the original dataset as our training set and the MTG-Giantsteps dataset as our test set.
For genre classification, we use the fault-filtered GTZAN dataset \cite{tzanetakis2002musical,sturm2013gtzan}. We use the VocalSet dataset \cite{wilkins2018vocalset} for two additional tasks: singer identification and technique classification. Finally, we regress Arousal (A) and Valence (V) on EmoMusic \cite{soleymani20131000} as a downstream evaluation task only, with the same train-test split as \cite{yuan2024marble}.

\subsection{Model input, augmentation chain}
\label{subsec: augmentation}

As in \cite{spijkervetContrastiveLearningMusical2021, vasquezTAILEDUNETMULTISCALE2022}, we crop 2.7 second segments of mono 22050kHz audio as input to the encoders, SampleCNN \cite{lee2018samplecnn} or TUNe+ \cite{vasquezTAILEDUNETMULTISCALE2022}. We sample and augment 2 adjacent non-overlapping segments as positives. The dimensions of the encoders and the 2-layer ReLU-nonlinear projection head are $d_E = 512$ and $d_g = 128$. We implement a stochastic augmentation chain similar to CLMR \cite{spijkervetContrastiveLearningMusical2021}, TUNe \cite{vasquezTAILEDUNETMULTISCALE2022}, and \cite{zhaoS3TSelfSupervisedPretraining2022}. In order, we apply (Table \ref{Tab: augmentations}):

\begin{table}[h]
\centering
\resizebox{\linewidth}{!}{%
\begin{tabular}{lllll}
\textbf{Augmentation} & \textbf{probability} & \textbf{parameter} & \textbf{Min/Max} & \textbf{unit} \\ \hline
Gain & 0.4 & Gain & -15$^\ddagger$ / 5$^\dagger$  & dB \\ \grayline
Polarity inv. & 0.6 & - & - & - \\ \grayline
Colored Noise & 0.6 & Signal/noise ratio & 3$^\ddagger$ / 30$^\ddagger$  & dB \\
 &  & Spectral decay & -2$^\ddagger$ 
/ 2$^\dagger$ & dB/octave \\ \grayline
\textit{Filtering} & (One of)  &  &  &  \\ 
Low pass & 0.3 & Cutoff & 0.15$^\ddagger$ / 7$^\ddagger$ & kHz \\
High pass & 0.3 & Cutoff & 0.2$^\dagger$ / 2.4$^\dagger$ & kHz \\
Band pass & 0.3 & center frequency & 0.2 / 4 $^\dagger$ & kHz \\
 &  &  Bandwidth fraction & 0.5$^\dagger$ / 2 & - \\
Band cut & 0.3 & center frequency & 0.2 / 4$^\dagger$ & kHz \\
 &  &  Bandwidth fraction & 0.5$^\dagger$ / 2 & - \\ \grayline
Pitch shifting & 0.6 & transpose & -4$^\ddagger$ / 4$^\dagger$ & semitones \\ \grayline
Delay & 0.6 & reflection time & 100$^\ddagger$ / 500& ms \\
 &  & reflections & 1$^\dagger$ / 3$^\dagger$ & - \\
 &  & attenuation & -6$^\dagger$ / -3$^\dagger$ & dB/reflection \\
 &  & wet/dry factor & 0.25$^\dagger$ / 1 & - \\ \hline
\end{tabular}%
}
\caption{Training augmentation chain. Only one amongst the four frequency filters is applied at once. Ranges denoted with $\dagger$ (resp. $\ddagger$) are subject to increasing (resp decreasing) in Subsection \ref{Subsec : robustness}}
\label{Tab: augmentations}
\end{table}

\subsection{Training and evaluation details}

For our baseline models, we adopt a training setup similar to TUNe \cite{vasquezTAILEDUNETMULTISCALE2022} and CLMR \cite{spijkervetContrastiveLearningMusical2021}. Models are trained for 200k steps on semi-supervised batches sampled from MagnaTagATune as labeled data and FMA-Medium as unlabeled data \cite{law2009evaluation, fma_dataset} using the Pytorch Adam optimiser with a learning rate of $1e^{-4}$. For ablation and variation studies, we train our models for 50k steps unless otherwise stated. All models are trained with $\tau=0.1$ with a non-augmented batch size of 96 on a single RTX A5000 GPU unless otherwise specified. We report steps instead of epochs to standardise the amount of data seen during training. 

To evaluate pretrained models, we freeze the encoder and discard the projection head. Frozen representations are fed into a 2-layer ReLU-nonlinear MLP for probing on downstream tasks. For probing, we use the Adam optimizer with a learning rate $0.0003$ and an early stopping mechanism conditioned on validation loss. For automatic tagging tasks, we report area under receiver-operator curve (AUROC) and mean Average Precision (AP). For classification tasks, we report top-1 accuracy except for key classification: the metric for this task is a weighted score taking into account reasonable errors \cite{yuan2024marble} - We use the \texttt{mir\_eval} \cite{raffel2014mir_eval} implementation for evaluation . For emotion regression we report $R^2$ values between predicted and actual values.


\section{Results}
\subsection{Automatic tagging with semi-supervised contrastive learning}
\label{mainresults}

\begin{table}
\centering
\resizebox{\linewidth}{!}{%
\begin{tabular}{lcccccc}
\cline{4-7}
 &
   &
   &
  \multicolumn{2}{c}{AUROC $\uparrow$}& \multicolumn{2}{c}{AP $\uparrow$}\\ \hline
\multicolumn{1}{l}{\textit{Ours}}&
  \textbf{$b_s = p_s$} & &
  \textbf{SampleCNN } &
  \textbf{TUNe+ }  & $\ddagger$ & $\bigstar$  \\ 
  
 &&&($\ddagger$)&($\bigstar$)&& \\
  \hline
\multicolumn{1}{l}{Self-Supervised}   & 0 &   & 88.8   &   88.9  & 41.6 & 41.6 \\ \grayline
\multicolumn{1}{l}{\multirow{6}{*}{Semi-Supervised}} &  0.05 &  &  89.4 &  89.4  & 42.5 & 42.1\\
\multicolumn{1}{l}{} & 0.1  &   & 89.5  &  89.4  & 42.2 & 42.2\\  
\multicolumn{1}{l}{}&  0.25  &   & 89.5 & 89.4 & 42.5 & 42.8\\
\multicolumn{1}{l}{}& 0.5  & & 89.7 & 89.5 & 42.9 & 43.3\\
    
\multicolumn{1}{l}{}  &  0.75  &    & 89.9 &  89.8  & 43.3& 43.5\\

\multicolumn{1}{l}{}   &  0.5/1  &    & 89.8 &  89.8  & 43.1& 43.0\\ \grayline
\multicolumn{1}{l}{\multirow{1}{*}{Supervised}} & \multirow{1}{*}{1} &  & \textbf{90.3} & 90.1 & 44.3 & \textbf{44.6} \\ \hline

\multicolumn{1}{l}{\textit{Literature}} &
   &
    &  &
    & &\\ \hline

    \multicolumn{1}{l}{SampleCNN \cite{spijkervetContrastiveLearningMusical2021}} &
  -  &
   &
  \multicolumn{2}{c}{89.3$^*$  (88.6$^\dagger$  \cite{spijkervetContrastiveLearningMusical2021})} &
  \multicolumn{2}{c}{41.2$^*$ (34.4$^\dagger$ \cite{spijkervetContrastiveLearningMusical2021})}\\ 
  \multicolumn{1}{l}{CLMR$_{\text{FMA}}$ \cite{spijkervetContrastiveLearningMusical2021}} &
  -  &
   &
  \multicolumn{2}{c}{86.6$^\dagger$} &
  \multicolumn{2}{c}{31.2$^\dagger$}\\ 
  
\multicolumn{1}{l}{TUNe+ \cite{vasquezTAILEDUNETMULTISCALE2022}}  &
  -  &
   &
  \multicolumn{2}{c}{89.2$^\dagger$} &
  \multicolumn{2}{c}{36.6$^\dagger$}\\
  \multicolumn{1}{l}{MERT \cite{liMERTAcousticMusic2023}} &
  -  &
   &
  \multicolumn{2}{c}{91.0$^\dagger$} &
  \multicolumn{2}{c}{39.3$^\dagger$}\\
  \hline
\end{tabular}
}
\caption{Performance on automatic tagging. Results denoted by $\dagger$ are reported in their original paper. In our experiment, we constrain $p_s = b_s$ except for one run where $p_s=1, b_s=0.5$. We trained our own end-to-end supervised SampleCNN with the same compute budget as SemiSupCon and report results with *.}
\label{table : preliminary results semisupcon}
\end{table}

\begin{figure}[h!]
    \centering
    \includegraphics[width = \linewidth]{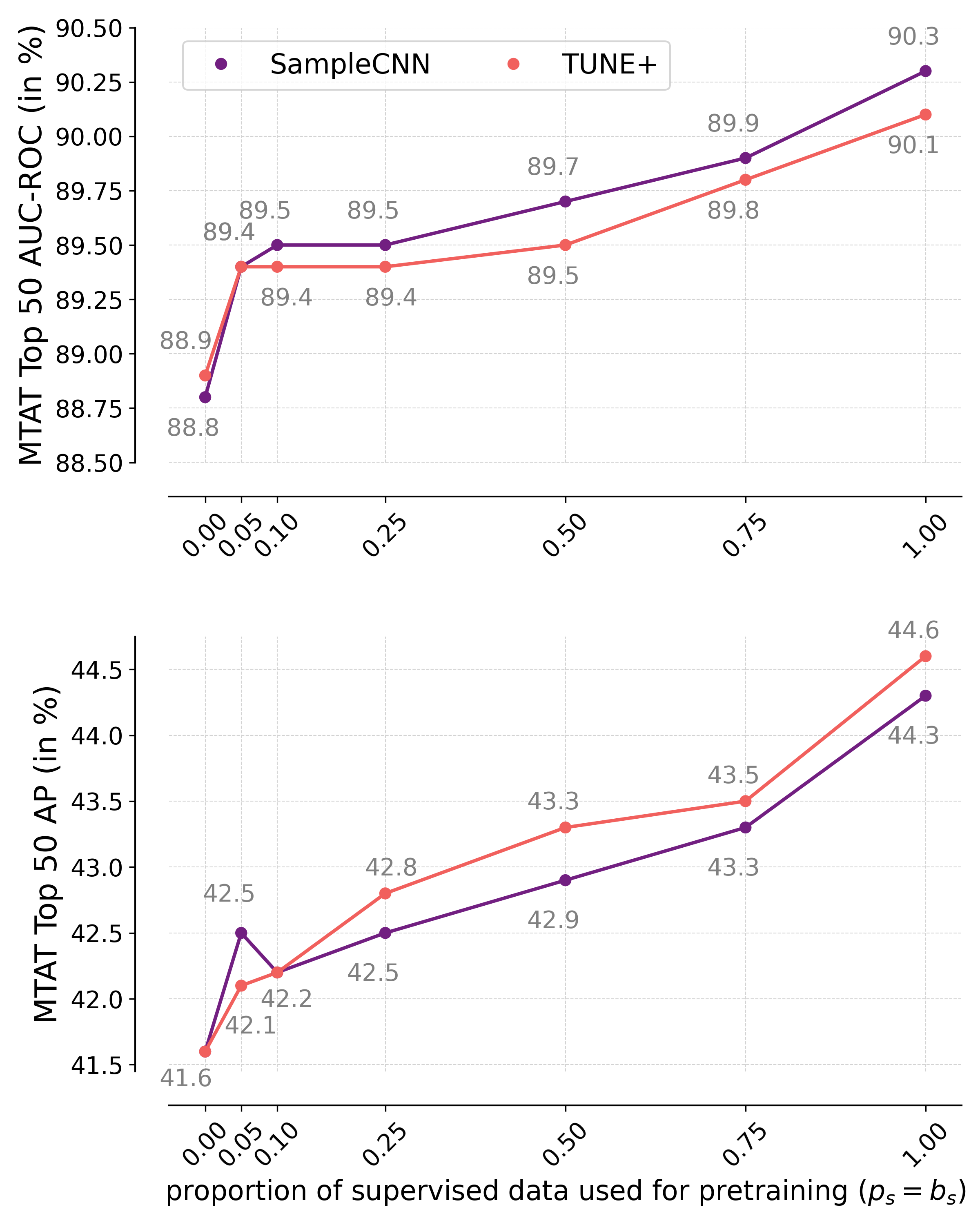}
    \caption{Evolution of AUROC and AP on MTAT probing with proportion of supervised MTAT data used for training.}
    \label{fig: main results}
\end{figure}

\vspace{-5pt}



We train a self-supervised baseline, a supervised contrastive baseline with and without augmentations, an end-to-end supervised baseline using the sampleCNN architecture, and five variants of our semi-supervised approach with different proportions of labeled data ($p_s \in [0.01,0.05, 0.1, 0.2, 0.5]$) for Automatic Tagging on MTAT. MTAT labels augment the contrastive matrix $\textbf{M}$ with positives in the case of supervised or semi-supervised pretraining. We vary the in-batch and global proportion of supervised data $b_s$ and $p_s$ simultaneously. We report results on the same task in the literature in Table \ref{table : preliminary results semisupcon} for comparable datasets and training scales.

When trained for 200k steps, the supervised contrastive model is competitive with larger self-supervised approaches. Furthermore, it outperforms both our implementation and the results claimed in CLMR for self-supervised contrastive and end-to-end supervised models. Figure \ref{fig: main results} shows the influence of $p_s = b_s$ on AUROC and AP. As the proportion of supervised data increases, so does the performance on the downstream evaluation. Including only 5\% of labeled data leads to an increase in performance from 88.8 to 89.4 in AUROC. For our experiment with $p_s = 1$ and $b_s = 0.5$, both architectures perform worse than $p_s = b_s = 0.75$, as the model has seen 100k steps of supervised data versus 150k.


\begin{table*}[t]
\centering
\resizebox{.95\textwidth}{!}{%
\begin{tabular}{llcccccccccccccc}
\hline
\multicolumn{2}{l}{Target Dataset} & \multicolumn{2}{c}{MTAT} & \multicolumn{4}{c}{Jamendo} & \multicolumn{2}{c}{NSynth} & \multicolumn{1}{c}{Giantsteps} & \multicolumn{1}{c}{GTZAN} & \multicolumn{2}{c}{VocalSet} & \multicolumn{1}{c}{MedleyDB} & \multicolumn{1}{c}{Emo} \\
\multicolumn{2}{l}{Subset} & \multicolumn{1}{c}{50} & \multicolumn{1}{c}{All} & \multicolumn{1}{c}{50} & \multicolumn{1}{c}{Genre} & \multicolumn{1}{c}{Mood} & \multicolumn{1}{c}{Inst.} & \multicolumn{1}{c}{Pitch} & \multicolumn{1}{c}{Inst.} & \multicolumn{1}{c}{Key} & \multicolumn{1}{c}{Genre} & \multicolumn{1}{c}{Tech.} & \multicolumn{1}{c}{Singer} & \multicolumn{1}{c}{Inst.} & \multicolumn{1}{c}{A/V} \\ \hline
\multicolumn{2}{l|}{Metrics} & \multicolumn{6}{c|}{AUROC} & \multicolumn{2}{c|}{Acc.} & \multicolumn{1}{c|}{Acc$_w$} & \multicolumn{4}{c|}{Acc.} & \multicolumn{1}{c}{$R^2_V$ / $R^2_A$} \\ \hline
\multicolumn{2}{l}{Self-Supervised} &  &  &  &  &  &    &  &  &  &  &  &  &  &  \\
FMA  & & 88.4  & 86.2  & 80.1 &  83.3& 74.0 & 71.6 & 36.8 & 51.7  &13.5 & 65.5 & 53.8 & 71.1 &  56.5 & 46.7/71.5   \\ \hline
\multicolumn{2}{l}{Semi-Supervised $b_s = 0.5$} &  &  &  &  &  &  &  &  &  &  &  &  &  &  \\
\multirow{2}{*}{MTAT} & 50  & \textbf{89.3} & 86.8 & 80.0 &  83.4& 73.8  & 73.3 & 34.5 &  46.9 &11.3 & 65.5 & 53.2  & 70.0 & 67.3 &  44.3/65.9 \\
 & All & 89.1 &\textbf{ 87.5} & 80.3 & 83.2 & 74.1 & 73.0 &  34.0 & 51.0  & 14.9 & 68.2& 52.4 & 72.9 &  \textbf{72.8} & 41.6/76.2 \\ \grayline
\multirow{4}{*}{Jamendo} & 50  & 88.6 & 86.6 & \textbf{81.5} & 83.4 & 74.6 & 72.5 & 33.8 & 50.0 & 14.7 &  \textbf{74.1} & 52.1 & 71.7 & 62.0 & 50.1/\underline{\textbf{77.9}} \\
 & Genre  & 88.6 & 86.3 & 80.5 &  \textbf{84.0}& 74.6 & 71.5 & 33.4 & 50.2 & 14.6 & 72.8 & 52.0 & 74.6 & 66.3 & 48.2/70.3 \\
 & Mood  & 88.3 & 86.6 & 81.0 & 83.0 & \textbf{74.7} & 72.3 & 38.2 & 47.7  & 14.9 & 71.3 & 53.5  & 71.4 & 60.9 & 48.0/73.0 \\
 & Instrument  & 88.4 & 86.3 & 80.8 & 83.1 & 74.0 & 71.6 & 37.2 & 52.5  & 14.9 & 69.3 & 54.5 & 67.9 & 63.0 & \textbf{52.4}/70.6 \\ \grayline
NSynth & Pitch$^\dagger$ & 88.3 & 86.3 & 79.7 &  82.6& 73.5 & 72.0 & \textbf{79.0} & 48.6   & 20.1 & 65.5 & 56.9 & 75.6 & 64.1 & 37.5/66.6 \\
 & Inst. & 88.6 & 85.7 & 79.6 &  82.7& 73.3 & 71.7 & 26.6 & \textbf{ 59.6} & 16.0 &  67.2 & 57.3 & 72.3 & 66.3 & 40.3/75.0 \\ \grayline
GiantSteps & Key$^\dagger$  & 87.7 & 85.0 & 79.0 &  82.1& 73.0 & 70.5 & 50.8 & 51.3 & \textbf{22.3 }& 69.3 & 54.1 & 71.4 & 61.2 & 39.6/63.6 \\ \grayline
GTZAN & Genre  & 88.8 & 86.8 & 80.9 & 83.9 & 74.1 & 71.5 & 38.6 & 46.9   & 16.3 & 74.0 & 53.4 & 71.7  & 66.3 & 28.7/56.4 \\ \grayline
\multirow{2}{*}{VocalSet} & Technique & 88.7 & 86.7 & 79.6  & 82.5  & 73.3 &  71.0 & 46.0 & 53.5 & 12.1 & 63.5 &  \textbf{63.0} & 77.8 & 67.3 & 41.5/70.1 \\
 & Singer & 88.9 & 86.2 &  80.1& 82.6 &  73.6 & 72.8 &  45.2 & 52.4  & 15.3 & 67.2 & 54.0 & \textbf{87.1} & 69.6 & 54.3/74.6 \\ \grayline
MedleyDB & Instrument & 88.6 & 87.0 & 80.2  & 82.6 & 73.6 & \textbf{73.8} &  32.0  & 48.8  & 13.2 & 62.1 & 58.6 & 74.3 & 62.0 & 41.6/74.8 \\ \hline
\multirow{2}{*}{\textcolor{gray}{ SOTA}} &  & \textcolor{gray}{92.7 } & \textcolor{gray}{95.4} & \textcolor{gray}{84.3}  & \textcolor{gray}{88.0} & \textcolor{gray}{78.6}  & \textcolor{gray}{78.8} & \textcolor{gray}{94.4}  & \textcolor{gray}{78.2} & \textcolor{gray}{74.3}  & \textcolor{gray}{86.9}   & \textcolor{gray}{76.9} & \textcolor{gray}{87.5}& \textcolor{gray}{-} & \textcolor{gray}{61.7/76.3}  \\
&  & \cite{wonFoundationModelMusic2023} &\cite{huangMuLanJointEmbedding2022} &\cite{mccallumSupervisedUnsupervisedLearning2022} &  \cite{castellonCodifiedAudioLanguage2021}  &   \cite{mccallumSupervisedUnsupervisedLearning2022} &  \cite{alonso-jimenezMUSICREPRESENTATIONLEARNING}& \cite{liMERTAcousticMusic2023} & \cite{wangLearningUniversalAudio2022} &  \cite{korzeniowski2017end} &   \cite{alonso2024leveraging} & \cite{liMERTAcousticMusic2023,yuan2024marble} &  \cite{liMERTAcousticMusic2023,yuan2024marble}&  & \cite{castellonCodifiedAudioLanguage2021,liMERTAcousticMusic2023}

\\ \grayline
\textcolor{gray}{ CLMR} \cite{yuan2024marble} & & \textcolor{gray}{89.5} & & \textcolor{gray}{81.3} & \textcolor{gray}{84.6} & \textcolor{gray}{73.5} & \textcolor{gray}{73.5} & \textcolor{gray}{47.0} &  \textcolor{gray}{67.9} & \textcolor{gray}{14.8} & \textcolor{gray}{65.2} & \textcolor{gray}{58.1} & \textcolor{gray}{49.9} & \textcolor{gray}{-} & \textcolor{gray}{44.4/70.3} \\
\hline
\end{tabular}%
}
\caption{Results for cross-task evaluation. Models are trained for 50k steps on FMA \cite{fma_dataset} as the self-supervised dataset and support supervised datasets (rows), and evaluated on target datasets (columns). Giantsteps$^\dagger$, NSynth$^\dagger$ are trained without pitch shifting augmentation. Results in bold are the best results obtained for evaluation on a target dataset. SOTA results are included for illustration purposes, but do not necessarily leverage comparable methodologies.}
\label{tab : cross-dataset}
\end{table*}

\subsection{Influence of pretraining labeled dataset}
\label{subsection: cross-dataset}

In this experiment, we pre-train multiple semi-supervised models using datasets described in Section \ref{datasets} as support labeled data and FMA as unlabeled data - one model per dataset, each for 50000 steps. We then freeze all models and train shallow MLP probes on all downstream tasks for each model. We train a self-supervised baseline for comparison. Semi-supervised approaches are trained with $b_s = 0.5$ and $p_s = 1$. Table \ref{tab : cross-dataset} shows these results. 


Semi-supervised training on the target dataset always surpasses the self-supervised baseline by a significant margin when evaluating on the same dataset - with minimal loss of performance on other downstream tasks. 


Some complementary tasks improve performance on other downstream datasets, proving semi-supervised contrastive learning a viable transfer learning strategy. Expectedly, training on genre tagging data increases out-of-domain performance on genre classification, instrument tagging on instrument classification, etc. Training on mood data from MTG-Jamendo provides a performance boost on emotion regression. A notable example is the improvement in performance on NSynth pitch when training with key data as support labeled data and \textit{vice versa}. This demonstrates an improvement in the understanding of \emph{pitch} by the model on tasks which are musically related but not trivial transfer learning instances. Most importantly, this occurs without performance loss on general music understanding, \textit{i.e.} automatic tagging. Other musically grounded examples are pitch pretraining improving instrument classification performance and instrument pretraining improving emotion regression performance.



\subsection{Robustness to in-domain data corruption}
\label{Subsec : robustness}

\begin{figure}
    \includegraphics[width=\linewidth]{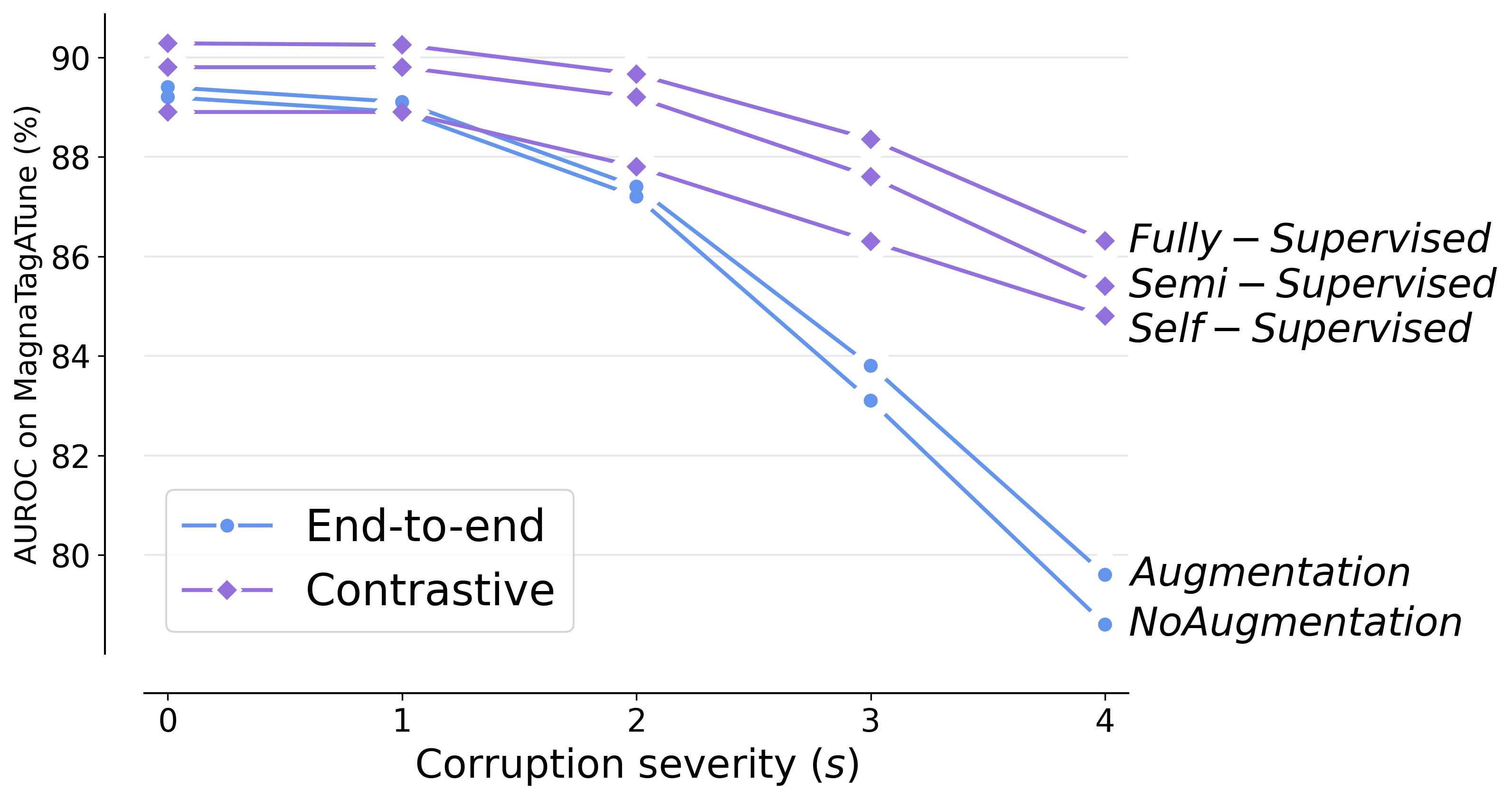}
    \caption{Effect of corruption severity on downstream performance. Contrastive models are more robust than Cross-entropy trained models.}
    \label{fig:robustness}
\end{figure}

In this section, we evaluate the robustness of our semi-supervised, supervised, and self-supervised contrastive approaches to audio corruptions compared to the end-to-end baseline. We train the probing head without augmentation until convergence and evaluate the model \emph{with} augmentations applied. We design different severity degrees of our augmentation chain (See Subsection \ref{subsec: augmentation}) by applying a modifier to the application probabilities: for severity $s \in [0,1...4]$, we scale probabilities of application of each augmentation by $s/2$ such that $s=2$ is the chain applied during training. We sensibly multiply or divide the min and max values of each augmentation hyperparameter (see Table \ref{Tab: augmentations}) by $s/2$. We then evaluate all models with these augmentation chains on MagnaTagATune. The results are shown in Figure \ref{fig:robustness}. Contrastive approaches are more robust to in-domain corruption than end-to-end approaches - hypothetically because we train contrastive models to be invariant to such transformations through the augmentation chain - which is not an objective of the end-to-end supervised approach.

\vspace{-5pt}

\subsection{Multilabel positive mining strategy}
\label{ablation: strategy}

In this experiment, we test multiple label-based positive mining strategies. First by varying the number of common labels for mining positives - i.e. $C \in \{1,2,4,6\}$. Further, we explore the ``semantic weighing'' strategy described in Section \ref{sec: SMSL_sup_signals}, in which the target similarity between two tracks is weighed by the number of common labels and the total number of labels. We test these strategies on both semi-supervised and supervised contrastive models. Results are reported in Table \ref{tab: strategy}.

\begin{table}[h]
\centering
\resizebox{.85\linewidth}{!}{%
\begin{tabular}{lllll}
\hline
Positive strategy & \multicolumn{2}{c}{Supervised} & \multicolumn{2}{c}{Semi-Supervised} \\ \hline
Class criterion & \multicolumn{1}{c}{AUROC} & \multicolumn{1}{c}{AP} & \multicolumn{1}{c}{AUROC} & \multicolumn{1}{c}{AP} \\ \hline
$C=1$ & 90.1 & 44.2 &  \textbf{89.3}&  41.3\\
$C=2$ & 90.1 & 43.9 &  89.0&  \textbf{41.6}\\
$C=4$ & 89.3 & 42.8 &  89.0&   41.3\\
$C=6$ & 88.9 & 42.3 &  89.0& 41.5\\ \hline
Weighing & \textbf{90.6} & \textbf{45.3} & 88.9& \textbf{41.6}\\ \hline
\end{tabular}%
}
\caption{Multilabel positive mining strategy as described in Section \ref{sec: SMSL_sup_signals}.}
\label{tab: strategy}
\end{table}

For supervised approaches, the continuous target produced by semantic weighing produces the best results, on par with 4x training steps with a criterion $C=1$ (as shown in Table \ref{table : preliminary results semisupcon}). In the supervised case, as the criterion increases, performance deteriorates. We hypothesise that this could be because it is an \emph{easier} task for the model to discern that two tracks with many common tags are similar (higher $C$), as they likely share many attributes, therefore providing a weaker training signal. Understanding what links two tracks from a single tag is more challenging and appears to yield more robust representations. The continuous ``relative similarity'' target created by the weighing strategy is a more nuanced task and appears to be a stronger supervision signal. This guides the model towards more robust representations, which explains the higher performance. In the semi-supervised case, we speculate that the binary self-supervision signal overpowers the continuous target as a less nuanced objective with harsher penalties for failure. These penalties could overpower softer penalties from the continuous target in the loss, preventing optimal convergence. Future work should focus on understanding and reconciling these aspects of the semi-supervised approach to leverage other continuous signals.


\vspace{-5pt}
\subsection{Qualitative analysis}
\label{qual}

\begin{figure}[h!]
    \centering
    \begin{subfigure}[b]{\linewidth}
        \centering
        \includegraphics[clip,width=\textwidth]{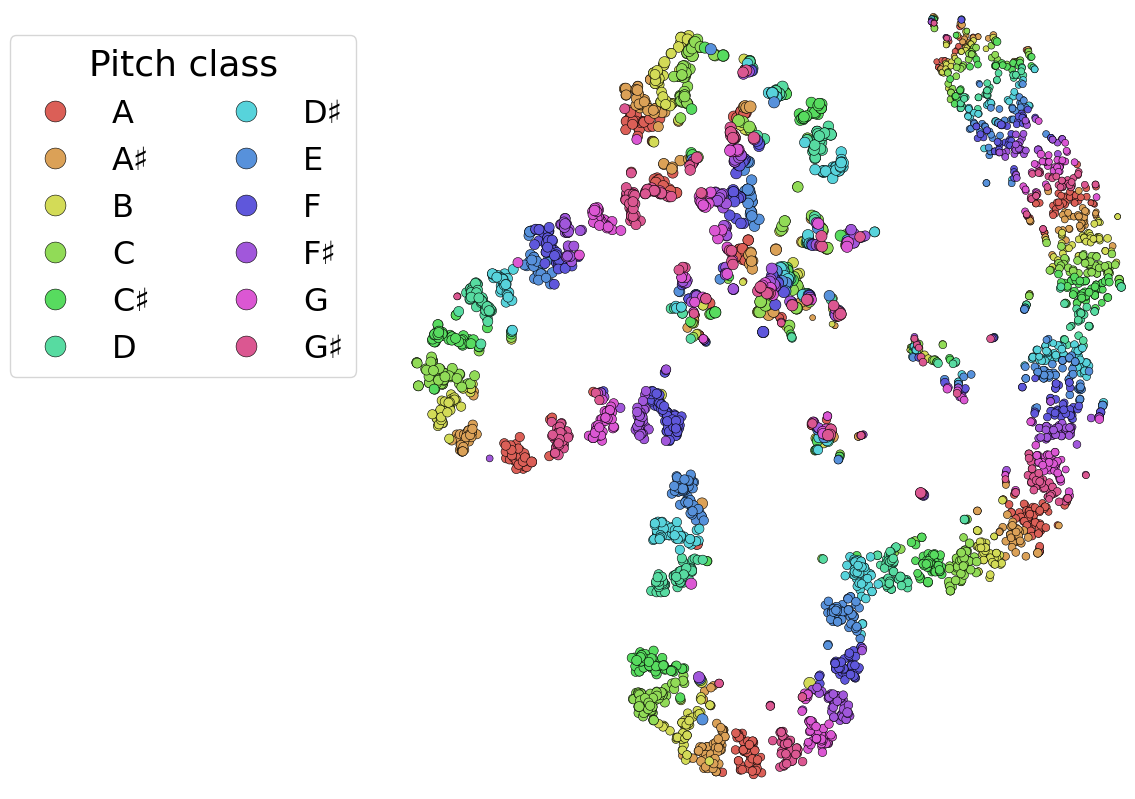}
        \caption{\centering Latent embeddings of the NSynth-pitch test set from a semi-supervised model trained on FMA+NSynth-pitch}
    
    \label{subfig : pitch}
    \end{subfigure}
    \begin{subfigure}[b]{0.78\linewidth}
        \centering
        \includegraphics[clip,width=\textwidth]{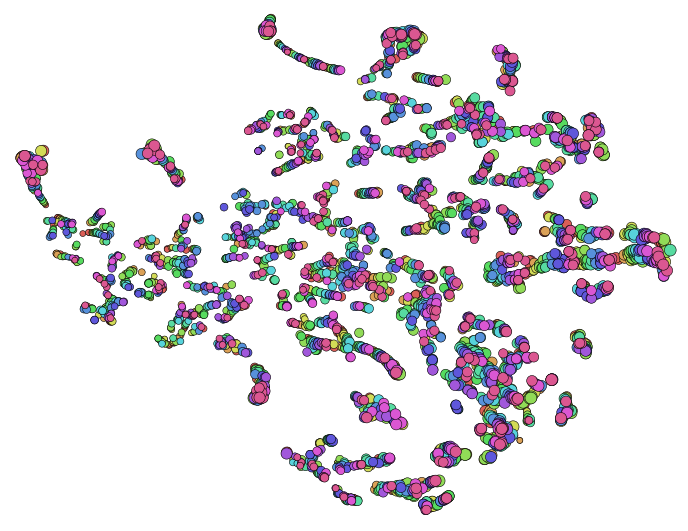}
        \caption{\centering Latent embeddings of the NSynth-pitch test set from a self-supervised model trained on FMA}
    \label{subfig : latent MTAT}
    \end{subfigure}
    \caption{Exploration of the NSynth pitch latent space. Octaves are denoted by size and pitch class by the color of the dot. Each dot is a full audio sample}
    \label{fig:viz_pitch}
\end{figure}

The results reported in Section \ref{subsection: cross-dataset} show that performance on downstream tasks improves when labels from a related task are used for model training, with minimal loss of performance on other tasks. We hypothesise that the internal latent representations are given structure relative to the supervision signal while maintaining the semantic structure given by the self-supervision signal. To illustrate this, we perform t-SNE dimension reduction on embeddings produced by the semi-supervised model from Table \ref{tab : cross-dataset} trained with NSynth (Figure \ref{subfig : pitch}) as support labeled data and fully self-supervised (Figure \ref{subfig : latent MTAT}) evaluated on the test set of NSynth-pitch.

In the set of Figures \ref{fig:viz_pitch}, the latent spaces for the NSynth test set produced by these two models are shown. When pretrained on NSynth-pitch, the latent space is highly organized. Separability by class is much clearer than when pretrained on FMA. We notice that several musical structures emerge in this latent space. Notably, octaves go from low to high clockwise. Pitches that are ``similar'' are close together, i.e., semitones and octaves of the same pitch class.


    
    

\vspace{-5pt}
\section{Conclusion and Future work}

We presented SemiSupCon, a simple method for leveraging both supervision and self-supervision signals in contrastive representation learning. By leveraging reduced amounts of labeled data during pretraining, SemiSupCon outperforms end-to-end comparable supervised baselines on downstream tasks. We find that SemiSupCon is more robust to data corruption at inference compared to end-to-end supervised methods. Additionally, SemiSupCon can utilize various supervision signals with minimal performance loss on out-of-domain tasks and achieve performance transfer on similar tasks. While performance gains might seem moderate on automatic tagging for instance, other downstream tasks show more distinct improvements. Furthermore, The contrastive objective can lead to explicitly structured latent spaces with emergent musical structures - enhancing the musical interpretability of latent spaces by design of the support supervision signal - i.e. labeling small amounts of data.

Future work will focus on exploring additional supervision signals and tasks such as perceptual metrics, tempo estimation, and chord estimation. Other avenues include leveraging the low-data proficiency of SemiSupCon for human-in-the-loop representation learning. The architecture of SemiSupCon being very flexible, it can be further adapted to multimodal approaches or hierarchical representation learning. A more comprehensive exploration of the influence of the proportion of labeled data and the exact effect of labels and contrastive matrix sparsity on downstream performance will also be undertaken.

\clearpage

\section{Acknowledgment}

This work is supported by the EPSRC UKRI Centre for Doctoral Training in Artificial Intelligence and
Music (EP/S022694/1) and Universal Music Group.

\bibliography{refs}

\clearpage

\import{./}{appendices.tex}

\end{document}

%% file: appendices.tex
\twocolumn[
\hrule
    \begin{center}
        {\Large \bfseries Appendix for the paper "Semi-Supervised Contrastive Learning of Musical Representations - preprint only}
    \end{center}
    \hrule
    \bigskip
    \bigskip
]

\section{Appendix}

\subsection{Training details for probing experiments}

In Section \ref{subsection: cross-dataset}, we evaluate representations learned through semi-supervised contrastive learning on a variety of datasets through shallow probing with MLP probes. To do so, we Freeze the contrastive backbone and train an MLP for the specified downstream task. Global training details are reported Section \ref{subsection: cross-dataset}. Here we go into probe architectures in more detail. We vary the number of hidden layers, dropout and weight decay for each dataset to account for overfitting. The values of hyperparameters were determined empirically through prelimiary runs. Probe architectures for different datasets as well as details on dataset scale are reported Table \ref{tab:datasets}:

\begin{table*}[h]
\centering
\resizebox{0.9\textwidth}{!}{%
\begin{tabular}{lllllllllll}
\toprule
\multicolumn{7}{c}{Dataset Details}                           & \multicolumn{1}{c}{} & \multicolumn{3}{c}{Probe details} \\ \cline{1-7} \cline{9-11} 
\multicolumn{1}{c}{Dataset} &
  Split &
  \multicolumn{1}{c}{\#classes} &
  \multicolumn{4}{c}{\#samples} &
  \multicolumn{1}{c}{} &
  \multicolumn{1}{c}{Probe layers output dims} &
  \multicolumn{1}{c}{Dropout} &
  \multicolumn{1}{c}{Weight Decay} \\ \cline{1-7} \cline{9-11} 
\multicolumn{1}{c}{} &
   &
  \multicolumn{1}{c}{} &
  \multicolumn{1}{c}{Train} &
  \multicolumn{1}{c}{Test} &
  \multicolumn{1}{c}{Validation} &
  \multicolumn{1}{c}{Total} &
  \multicolumn{1}{c}{} &
  \multicolumn{1}{c}{} &
  \multicolumn{1}{c}{} &
  \multicolumn{1}{c}{} \\ \cline{1-7} \cline{9-11} 
FMA       &           & N/A & N/A    & N/A   & N/A   & 25000  &                      & N/A          & N/A    & N/A       \\ \cline{1-7} \cline{9-11} 
MTAT      & All       & 288 & 18709  & 5329  & 1825  & 25863  &                      & 512, 288     & 0.1    & $1e-6$    \\
          & Top50     & 50  & 15250  & 4332  & 1529  & 21111  &                      & 512, 50      & 0.1    & $1e-6$    \\ \cline{1-7} \cline{9-11} 
Jamendo   & Top50     & 50  & 32136  & 11356 & 10888 & 54380  &                      & 512,50       & 0.1    & $1e-6$    \\
          & Mood      & 56  & 9949   & 4231  & 3802  & 17982  &                      & 512, 56      & 0.1    & $1e-6$    \\
          & Genre     & 87  & 32572  & 11479 & 11043 & 55094  &                      & 512, 87      & 0.1    & $1e-6$    \\
          & Instr.    & 40  & 14395  & 5115  & 5466  & 24976  &                      & 512, 40      & 0.1    & $1e-6$    \\ \cline{1-7} \cline{9-11} 
Nsynth    & Pitch     & 112 & 289205 & 4096  & 12678 & 305979 &                      & 512, 112     & 0      & $1e-6$    \\
          & Instr.    & 24  & 289205 & 4096  & 12678 & 305979 &                      & 512, 24      & 0.1    & $1e-6$    \\ \cline{1-7} \cline{9-11} 
Giansteps &           & 24  & 961    & 604   & 198   & 1763   &                      & 24           & 0.2    & $1e-5$    \\ \cline{1-7} \cline{9-11} 
GTZAN     &           & 10  & 443    & 290   & 197   & 930    &                      & 512, 10      & 0.1    & $1e-6$    \\ \cline{1-7} \cline{9-11} 
VocalSet  & Singer    & 20  & 2467   & 577   & 566   & 3610   &                      & 20           & 0.2    & $1e-6$    \\
          & Technique & 17  & 2282   & 894   & 403   & 3576   &                      & 17           & 0.2    & $1e-6$    \\ \cline{1-7} \cline{9-11} 
MedleyDB  &           & 20  & 659    & 137   & 117   & 913    &                      & 20           & 0.2    & $1e-6$    \\ \cline{1-7} \cline{9-11} 
EmoMusic  &           & 2*  & 504    & 125   & 115   & 744    &                      & 2            & 0.1    & $1e-5$    \\ \bottomrule
\end{tabular}%
}
\caption{Dataset and probe details for fine-tuning experiments. For Emomusic (*) 2 classes are reported but are in fact regression targets (A/V)}
\label{tab:datasets}
\end{table*}

\subsection{Additional qualitative analysis}

Here, we provide additional visualizations for qualitative analysis of the embeddings obtained through cross-dataset training. The embeddings are obtained from a model trained semi-supervisedly with FMA as the self-supervised dataset and Nsynth-pitch as the supervised dataset. Embeddings are obtained with the frozen model for the MagnaTagATune test set and t-SNE is applied to obtain visualizations similar to \ref{qual}.

\begin{figure}[h!]
    \centering
    \begin{subfigure}[b]{\linewidth}
        \centering
        \includegraphics[clip,width=\textwidth]{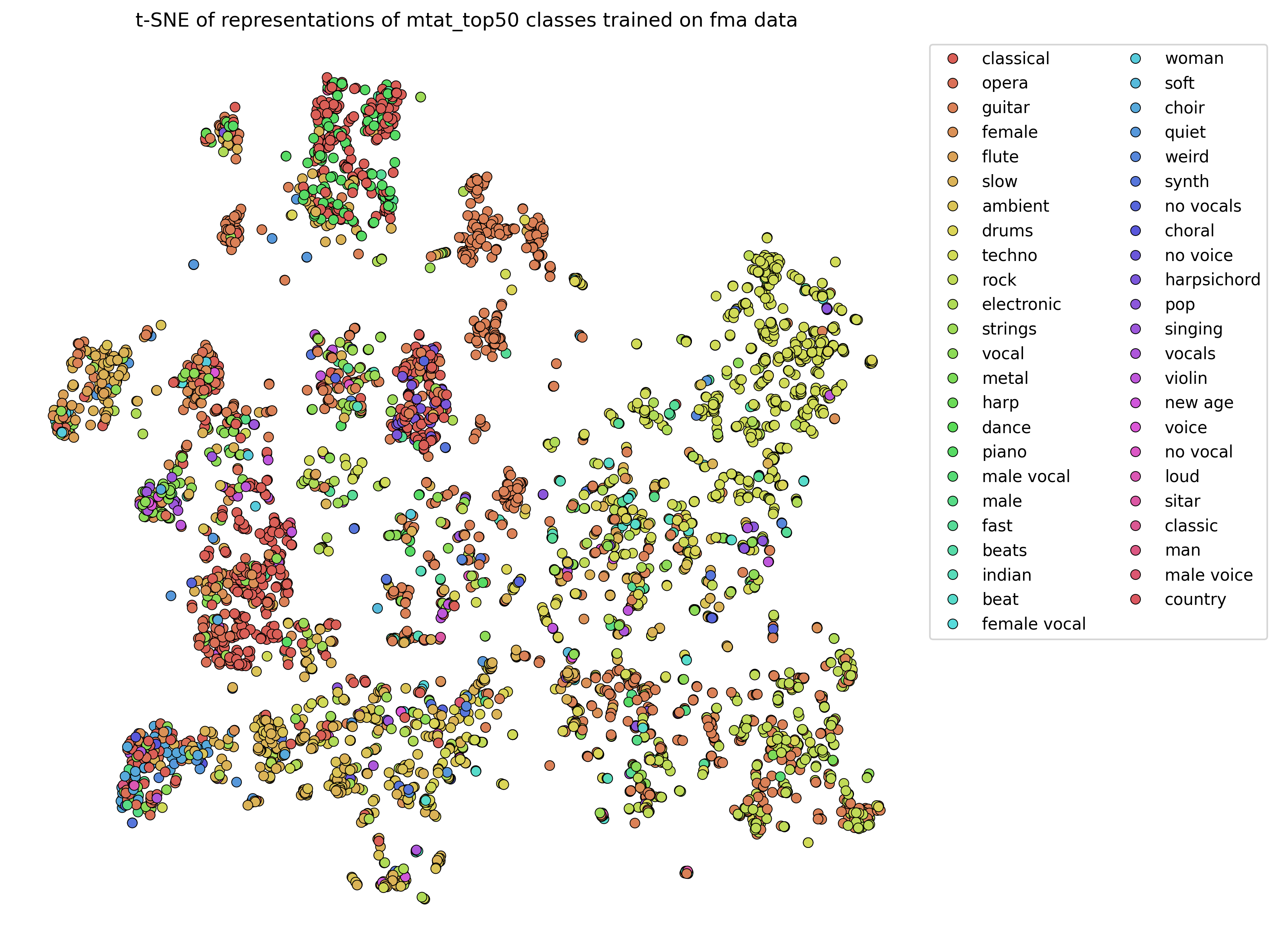}
        \caption{\centering Latent embeddings of the NSynth-pitch test set from a semi-supervised model trained on FMA+NSynth-pitch}
    
    \end{subfigure}
    \begin{subfigure}[b]{1\linewidth}
        \centering
        \includegraphics[clip,width=\textwidth]{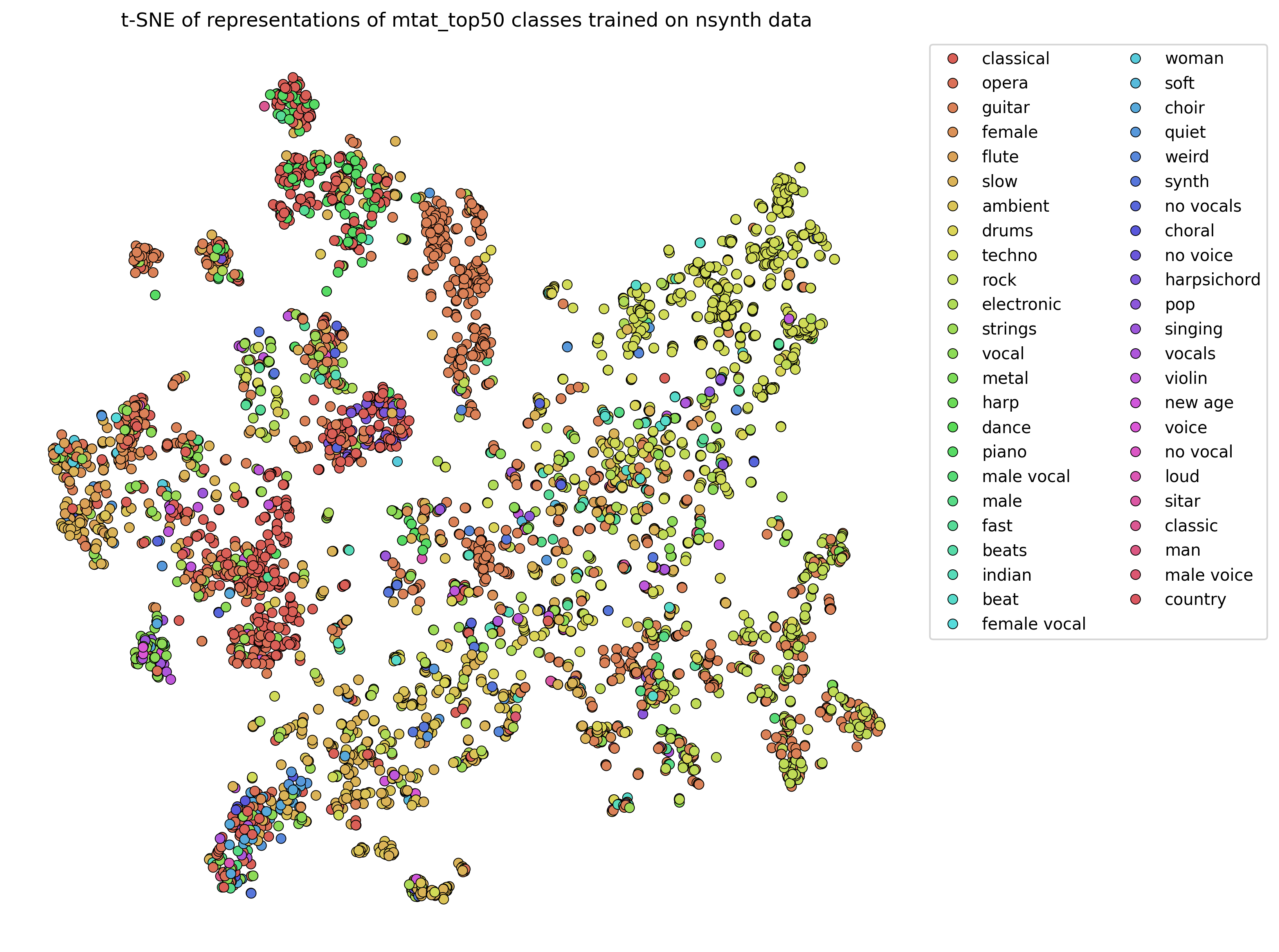}
        \caption{\centering Latent embeddings of the NSynth-pitch test set from a self-supervised model trained on FMA}
    \end{subfigure}
    \caption{Exploration of the NSynth pitch latent space. Octaves are denoted by size and pitch class by the color of the dot. Each dot is a full audio sample}
\end{figure}

Where training with Nsynth as support labeled data greatly enhanced the structure of pitch-related classes within the embedding space, we find that the structure of other datasets (in this case, automatic tagging) is not lost, and we even observe the same cluster structure, with some small details. This corroborates our findings in Section \ref{subsection: cross-dataset} that performance can be greatly improved on a dataset of interest without losing general performance and understanding on other tasks through semi-supervised contrastive learning, a key benefit of our method.

\subsection{Additional ablation studies}
\subsubsection{Individual variation of $p_s$ and $b_s$}
\label{ablation : psbs}
To investigate the influence of $p_s$ and $b_s$ on their own, we design two experiments : In one, $p_s$ the proportion of supervised data is fixed at $p_s = 1$ and $b_s \in [0.05,0.1,0.25,0.5,0.75]$ is varied. In another, $b_s$ is fixed at $b_s = 0.5$ (exactly half of each batch is supervised data) and $p_s \in [0.05,0.1,0.25,0.5,0.75]$ is varied. We use the SampleCNN architecture for all our experiments. All models are pretrained with MTAT as the support labeled data as well as the downstream task and FMA as the self-supervised dataset. To better relate the in-batch proportion of supervised data to a general property of the semi-supervised contrastive learning task, we define the semi-supervised and supervised contrastive matrix sparsities $s_{smsl}$ and $s_{sl}$:

\begin{equation*}
    s_{smsl} = \frac{1}{|I|} \sum\limits_i |P_A(i)| 
\end{equation*}

and 
\begin{equation*}
    s_{sl} = \frac{1}{b_s|I|} \sum\limits_i |P_s(i)| 
\end{equation*}

Which represent proportion of positives in the contrastive matrix. as this attribute of the system depends on many hyperparameters such as the task, the positive sampling strategy, the in-batch supervised proportion, and the number of augmentations, we report it in Table \ref{tab:psbs}. 

\begin{table}[h]
\centering
\resizebox{.45\textwidth}{!}{%
\begin{tabular}{cccccc}
\toprule
\multicolumn{1}{l}{$p_s$} & \multicolumn{1}{l}{$b_s$} & \multicolumn{1}{l}{$s_{sl}$ (\%)} & \multicolumn{1}{l}{$s_{smsl}$ (\%)} & \multicolumn{1}{l}{\textbf{AUROC}} & \multicolumn{1}{l}{\textbf{AP}} \\ \hline
1 & 0.05 & 87.5 & 99.5 & 88.4 & 40.6 \\
1 & 0.1 & 77.1 & 99.3 & 88.4 & 40.2 \\
1 & 0.25 & 61.5 & 97.2 & 88.2 & 40.4 \\
1 & 0.5 & 74.7 & 93.4 & 89.0 & 41.4 \\
1 & 0.75 & 77.3 & 87.1 & 89.3 & 42.0 \\ \hline
\multicolumn{1}{l}{} & \multicolumn{1}{l}{} & \multicolumn{1}{l}{} & \multicolumn{1}{l}{} & \multicolumn{1}{l}{} & \multicolumn{1}{l}{} \\ \hline
\multicolumn{1}{l}{$p_s$} & \multicolumn{1}{l}{$b_s$} & \multicolumn{1}{l}{$s_{sl}$ (\%)} & \multicolumn{1}{l}{$s_{smsl}$ (\%)} & \multicolumn{1}{l}{\textbf{AUROC}} & \multicolumn{1}{l}{\textbf{AP}} \\ \hline
0.05 & 0.5 & 69.7 & 92.2 & 88.4 & 40.4 \\
0.1 & 0.5 & 73.3 & 93.1 & 88.2 & 40.2 \\
0.25 & 0.5 & 74.4 & 93.3 & 89.0 & 41.0 \\
0.5 & 0.5 & 65.5 & 91.2 & 88.4 & 40.6 \\
0.75 & 0.5 & 77.1 & 94.0 & 88.6 & 40.9 \\ \bottomrule
\end{tabular}%
}
\caption{Individual variation of $p_s$ and $b_s$: varying $b_s$ leads to a clear improvement, while $p_s$ seems to lead to more fluctuations in performance.}
\label{tab:psbs}
\end{table}

We find that varying $b_s$ leads to a consistent increase in performance as $b_s$ increases, while the same does not hold true for $p_s$. This is reasonable : When increasing $b_s$ for the same amount of labeled data and training steps, the amount of labeled data the model sees during training increases, thus reinforcing the influence of labeled data and bringing the model closer to a fully-supervised contrastive setup. For varying $p_s$ however, it is not clear as to why the performance does not increase consistently. Perhaps the noisiness of the data here is a factor that explains why different subsets of labeled data (potentially with different distributions to the test set) seen during training would lead to fluctuations in downstream performance. This warrants further exploration to understand why increasing the diversity of labeled data does not consistently improve performance. 

\subsubsection{Influence of number of augmentations for same batch size}

Because the inclusion of more positives into the training pipeline through explicit labeling seems to improve the performance of the model significantly, a possible intuition following this observation is that including more positive samples in the contrastive matrix could lead to similar improvements. For this study, we kept the global batch size constant at 192 and vary the number of augmentations and the batch size without augmentations. We report results on MTAT for fully supervised, self-supervised, and for $p_s = 1$ and $b_s = 0.5$ for training.

\begin{table}[h]

    \centering
\resizebox{.45\textwidth}{!}{%
    \begin{tabular}{llllllllll}
 \toprule
 &  & & & \multicolumn{2}{c}{SSL} & \multicolumn{2}{c}{SL} & \multicolumn{2}{c}{Semi-SL} \\ \hline
$M$ & $B$ & $s_{sl}$ & \multicolumn{1}{c|}{$s_{smsl}$} & AUROC & \multicolumn{1}{c|}{AP} & AUROC & \multicolumn{1}{c|}{AP} & AUROC & AP \\
\hline 
2 & 96 & 74.7 & 93.5 & 88.4 & 40.1 & 90.1 & 44.1 & 89.3  & 41.0 \\
4 & 48 & 69.4 & 88.3 & 88.5 & \textbf{40.6} & 90.4 & 44.8 & 89.2 & \textbf{42.0} \\
8 & 24 & 60.1 & 82.7 & \textbf{88.6} & 40.3 & \textbf{90.5} & \textbf{44.9} & \textbf{89.6} & 41.9 \\
16 & 12 & 51.0 &76.1 & 88.1& 39.5 & 89.3 & 41.9 & 88.3 & 40.2 \\
32 & 6 & 40.1 & 73.1.4 & 87.4 & 38.4 & 88.9 & 40.7 & 87.4 & 38.4 \\ \toprule
\end{tabular}}
    \caption{Influence of the number of augmentations $M$ on downstream MTAT performance for SSL, SL, and SMSL models}
    \label{tab:augmentations}
\end{table}

Here, it is not fully clear how the number of augmentations improve performance. However, there seems to be a sweet spot for a given batch size of moderate number of augmentations. In this case, 8 augmentations for a batch size of 24 unaugmented samples seems to be benefecial to performance across the board for AUROC. AP benefits from 4-8 augmentations more than other numbers of augmentations. Evidently this result is dependent on the batch size as well as the augmentation strategy. In our case, we find that a moderate amount of augmentation-sampled positives yields stronger representations, contrary to the well-known \textit{a priori} that only the number of negatives for each sample in contrastive learning is an important factor for downstream performance.

\subsection{Additional adversarial robustness studies}.

We also test the robustness of contrastive and cross-entropy models when subjected to novel corruption transformations at inference time. We implement new augmentations with the torch-audiomentations library and the Spotify pedalboard library, Including \textbf{Chorus, Distortion, SpliceOut, Reverb, TimeStretch, Compression, and Bitcrush}.

\begin{figure}[h]
    \includegraphics[width=.5\textwidth]{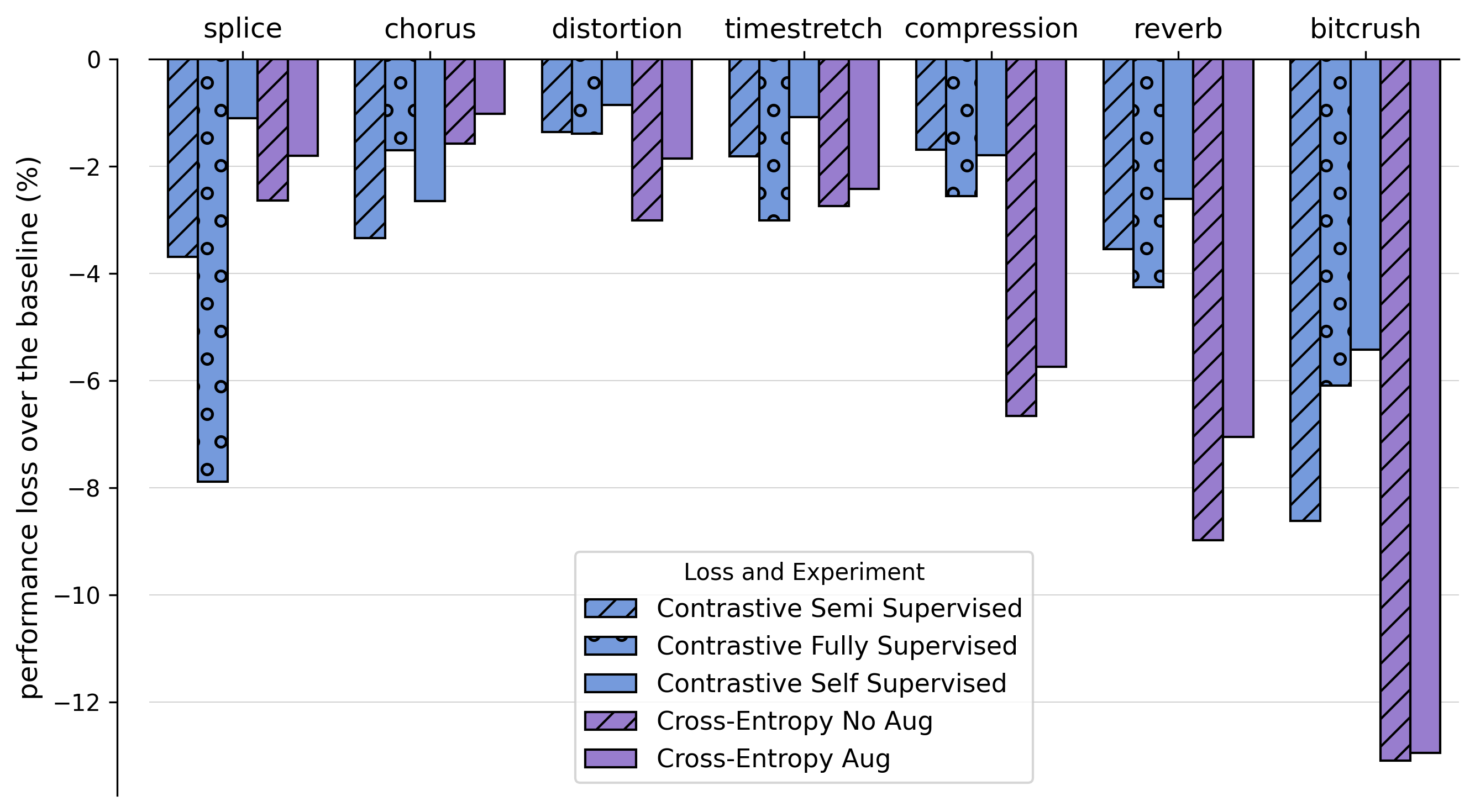}
    \caption{Effect of novel corruptions on downstream AUROC performence - relative performance variation when compared to uncorrupted baselines.}
    \label{fig:robustness2}
\end{figure}

In most cases, except the notable exception of SpliceOut and Chorus, contrastive approaches are more robust to out-of-domain corruption when compared to cross-entropy approaches.